\documentclass[conference]{IEEEtran}
\usepackage{color, soul}

\usepackage{colortbl}
\newcommand*\rot{\rotatebox{90}}

\ifCLASSINFOpdf
  \usepackage[pdftex]{graphicx}
  \usepackage{tikz}
\else
\fi
\begin{document}
%
\title{Sonification in Security Operations Centres: What do Security Practitioners Think?}

\author{\IEEEauthorblockN{Louise Axon, Bushra Alahmadi, Jason R. C. Nurse, Michael Goldsmith and Sadie Creese}
\IEEEauthorblockA{Department of Computer Science, University of Oxford\\
\{louise.axon, bushra.alahmadi, jason.nurse, michael.goldsmith, sadie.creese\}@cs.ox.ac.uk}}


%


\IEEEoverridecommandlockouts
\makeatletter\def\@IEEEpubidpullup{6.5\baselineskip}\makeatother
\IEEEpubid{\parbox{\columnwidth}{
    Workshop on Usable Security (USEC) 2018 \\
    18 February 2018, San Diego, CA, USA \\
    ISBN 1-891562-53-3 \\
    https://dx.doi.org/10.14722/usec.2018.23024 \\
    www.ndss-symposium.org
}
\hspace{\columnsep}\makebox[\columnwidth]{}}

\maketitle

\begin{abstract}
In Security Operations Centres (SOCs) security practitioners work using a range of tools to detect and mitigate malicious computer-network activity. Sonification, in which data is represented as sound, is said to have potential as an approach to addressing some of the unique challenges faced by SOCs. For example, sonification has been shown to enable peripheral monitoring of processes, which could aid practitioners multitasking in busy SOCs. The perspectives of security practitioners on incorporating sonification into their actual working environments have not yet been examined, however. The aim of this paper therefore is to address this gap by exploring attitudes to using sonification in SOCs. We report on the results of a study consisting of an online survey \textit{(N=20)} and interviews \textit{(N=21)} with security practitioners working in a range of different SOCs. Our contribution is a refined appreciation of the contexts in which sonification could aid in SOC working practice, and an understanding of the areas in which sonification may not be beneficial or may even be problematic. We also analyse the critical requirements for the design of sonification systems and their integration into the SOC setting. Our findings clarify insights into the potential benefits and challenges of introducing sonification to support work in this vital security-monitoring environment.
\end{abstract}


%

\section{Introduction}\label{sec:introduction}
The threats to the cybersecurity of today's organisations are numerous, vastly varied and constantly evolving. Security Operations Centres (SOCs) run within and on behalf of organisations, and are responsible for the security of networks and critical infrastructure. In SOCs, security practitioners work, often under high pressure \cite{sundaramurthy2015human}, interacting with a range of security tools to detect and prevent malicious activity. There is a requirement for monitoring tools for use in SOCs that are effective and meet the needs of security practitioners. In recent years, the incorporation of sonification, in which data is represented as sound, into SOCs has been considered. 

Sonification is defined as ``the use of non-speech audio to convey information'' \cite{kramer1999sonification}. The outputs of sonification systems are often referred to as ``sonified displays", or ``auditory displays". A body of research exists into the use of sonification for monitoring processes, exploring data, and alerting \cite{hermann2011sonification}. Based on existing research, the properties afforded by sonification align with some known requirements of SOCs. Articles exploring the sonification of network-security data indicate its promise as a technique for attack detection \cite{axon2017formalised, ballora2011songs, mancuso2015augmenting}, improved methods for which are critical to SOCs. Furthermore, sonification is an effective medium for peripheral monitoring of information as a non-primary task \cite{hildebrandt2016continuous}. This could be useful to busy practitioners in bustling SOCs. On the other hand, there are concerns about the fatigue and distraction that could be caused by sonification, which raise questions about its true utility in these dynamic environments.

Despite these potential benefits, there has to-date been no research exploring practitioners' perspectives on the contexts in which sonification could integrate into SOC workflow. It is therefore unclear how these practitioners regard the incorporation of sonification into SOCs. Understanding the needs of users, however, is crucial to incorporating new technologies into their working environment \cite{bevan2001international}. To address this gap, we consult practitioners working in SOCs, to explore their perspectives on incorporating sonification into this unique setting. As an initial stage in the user-centred design process \cite{maguire2002user}, our aim is to identify and refine contexts of use for sonification in SOCs, and analyse integration and design requirements.

This paper reports on the results of a study involving an online survey and semi-structured interviews. Firstly, we designed tentative use-cases for sonification in SOCs, using information gathered from existing literature and the responses of security practitioners in an online survey. We then carried out semi-structured interviews with security practitioners working in SOCs. In these interviews we presented participants with a network-packet sonification prototype we developed, in order to familiarise them with the concept of sonification. The proposed tentative use-cases were then explored, and participants' views on integration and design discussed. We thus refined contexts of use, discarding use-cases that were not considered to have promise, and analysed user needs with regard to integration and design \cite{gulliksen2003key}.

The paper makes the following contributions to the Human-Computer Interaction (HCI) and Usable Security domains:

\begin{itemize}
\item Identifies and refines the contexts in which sonification systems could improve working practice in SOCs.
\item Establishes an empirical understanding of the challenges of integrating sonification into the SOC setting.
\item Extracts design requirements for sonification tools that would be effective and usable for SOC practitioners.
\end{itemize}

Our findings can inform sonification interface development, and future studies into the use of sonification in SOCs.

\section{Background and Related Work}\label{sec:background}

We begin with an overview of the work of security practitioners in SOCs. We then review HCI studies on SOC work, and applications of sonification to network-security tasks.

\subsection{Security Operations Centres (SOCs)}

\begin{figure}[ht]\label{fig:soc_image}
\centering
\includegraphics[trim=0 700 0 250,clip,scale=0.075]{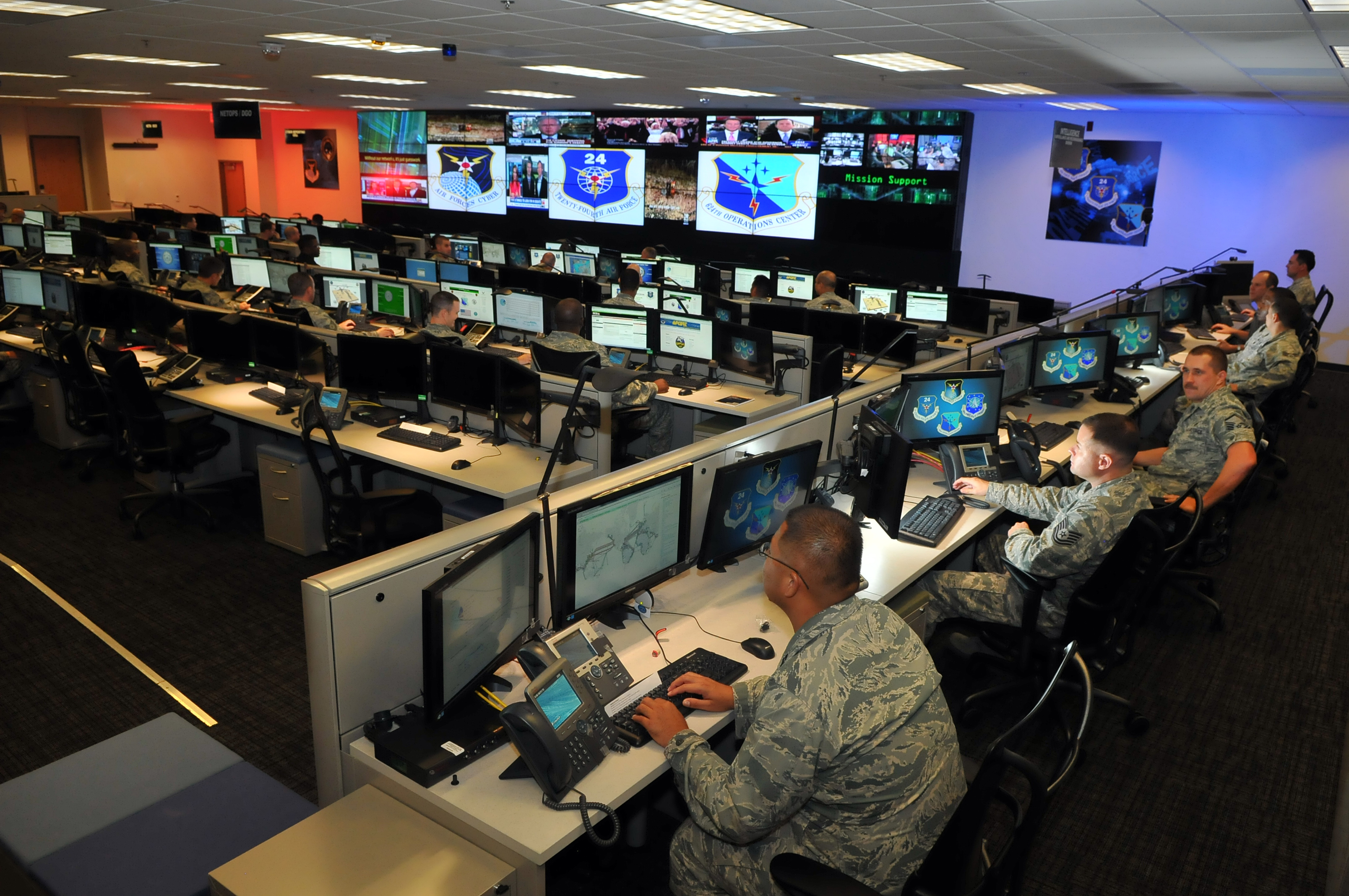}
\caption{A Security Operations Centre (SOC) (U. S. Air Force photo) \protect\cite{soc_photo}}
\end{figure}

The objective of a SOC is primarily to mitigate cybersecurity threats towards the organisations for which they are responsible \cite{sundaramurthy2017humans}. Internal SOCs are responsible for the organisations they are placed within, while multitenanted SOCs monitor network security on behalf of multiple client organisations. Figure \ref{fig:soc_image} is an example of a SOC, with security data presented to security practitioners on computer monitors. Practitioners are frequently required to work long shifts, including night shifts, looking at multiple screens for extended time periods \cite{sundaramurthy2014tale}. The resulting pressure and demanding nature of SOC work have been highlighted in HCI research \cite{sundaramurthy2015human, sundaramurthy2014tale}.

Security practitioners interact with automated security tools, such as signature- or anomaly-based intrusion-detection systems (IDSs), which produce security events. This data is often collated in integrated security incident and event management (SIEM) solutions \cite{sundaramurthy2014tale}. The role of security analysts can include preliminary detection, triage of events, and responding to customer tickets. Security engineers are also responsible for maintaining infrastructure and creating detection rules, for example \cite{sundaramurthy2014tale}. By ``security practitioner'', we denote a person who works in a SOC (an analyst, engineer, or manager).

It is important that humans are presented with network-security information, such that they are best able to detect anomalies that do not fit automated detection profiles, and to triage machine-based detection inaccuracies \cite{d2016cyber}. The provision of effective techniques for presenting security data to humans is important for SOCs, and an area of continuing academic research \cite{zhang2012survey}. Security visualizations and text-based interfaces present automated system output, as well as unparsed network packets, which can enable security practitioners to recognise anomalous activity \cite{botta2007towards}.

\subsection{HCI Studies in SOCs}

A number of HCI articles have focused on examining the work of security practitioners in SOCs, and the challenges faced. This has included interview-based research \cite{d2008real, werlinger2009integrated, werlinger2010preparation}, and ethnographic fieldwork \cite{d2005achieving, sundaramurthy2014tale, werlinger2008security}. Below, we reflect on some of the most pertinent to our research.

Sundaramurthy et al. conducted anthropological fieldwork in SOCs spanning 4 years. Students trained in anthropological methods were embedded in three different SOCs as security analysts \cite{sundaramurthy2015human, sundaramurthy2014tale, sundaramurthy2016turning, sundaramurthy2017humans}. Activity Theory was used to model SOC operations, and the successes and failures encountered in integrating new technologies into SOCs studied. The implications of the findings for improving SOC operations were described, including the need for useful new tools to be dynamic and constantly resolve emerging conflicts \cite{sundaramurthy2016turning}. Factors contributing to security analyst burnout, rates of which are consistently high, were modelled as a cycle linking factors concerning skills, empowerment, creativity and growth \cite{sundaramurthy2015human}. 

Werlinger et al. used interviews and participatory observation to identify the interactions of security practitioners \cite{werlinger2008security, werlinger2009integrated, werlinger2009security}. They found that the existing tools used were not sufficient to support complex security tasks, with the high number of false positives produced by IDSs highlighted  \cite{werlinger2009integrated}. In extended research, Werlinger et al. used semi-structured interviews to understand security incident response practices \cite{werlinger2010preparation}. Findings included a tendency for complication of incident diagnosis by usability issues with security tools, and by a need for practitioners to rely on their own knowledge.

D'Amico et al. investigated the workflow, decision processes, and tool use of security practitioners in SOCs using cognitive task analysis \cite{d2005achieving}. Cognitive challenges including the massive amounts of network data were identified. D'Amico et al. also explored the perspectives of security practitioners on the use of security visualizations in their work \cite{d2016cyber}. Findings indicated that visualizations could support data analysis. 

While HCI studies have identified approaches to improving SOC operations, approaches using sonification have not been explored. The use of sonification has been examined only insofar as its utility in network-security tasks has been assessed, in studies not specific to SOCs (reported in Section \ref{subsec:litrev_sonnw}). Incorporating sonification into SOCs has not, to our knowledge, been explored from an HCI perspective.

\subsection{Sonification for Network-Security Monitoring}\label{subsec:litrev_sonnw}

Prior work has applied sonification in security-monitoring tasks. Axon et al. surveyed existing articles \cite{axon2017formalised}, highlighting sonification systems designed for network-attack detection \cite{ballora2011songs, brownposter, gilfix2000peep, giot2012intention, mancuso2015augmenting, papadopoulos2004cyberseer, qi2007toward}. The utility of sonification for SOCs is proposed, based on the challenges SOCs face, and evidence of the potential benefits of sonification \cite{axon2017formalised}. Sonification can enable humans  either to identify a general change in status, without knowing exactly what changed, or to actually understand the meaning of the information represented.

Researchers have reported the ability to hear attacks using a range of mappings from network traffic features to parameters of sound \cite{ballora2011songs, qi2007toward}. Qi et al. mapped network-traffic parameters to sound, and stated that a range of attack scenarios were distinguishable \cite{qi2007toward}. Ballora et al. sonified network traffic with a view to aiding anomaly detection, and reported the ability to hear patterns associated with port-scanning and distributed denial-of-service (DDoS) attacks \cite{ballora2011songs}. Gilfix et al. detected unusual network conditions such as excessive traffic using a mapping from network traffic to natural sounds \cite{gilfix2000peep}. 

User studies have been carried using with sonification systems for network-security monitoring tasks. Gopinath sonified a range of security events in Snort IDS \cite{gopinath2004auralization}. Results indicated that sonification may increase user awareness in intrusion detection \cite{gopinath2004auralization}. Mancuso et al. assessed the detection of packets with particular characteristics by cyber operators using a sonification whilst searching through a packet capture. The particular sonification design used in that study, however, did not improve participants' packet-detection capabilities \cite{mancuso2015augmenting}. 

Kaczmarek et al. found that non-expert participants' failure rates in carrying out security-critical tasks were lower when auditory cues were played \cite{kaczmarek2015unattended}. Less complex stimuli improved performance, while more complex stimuli worsened it \cite{berg2017exploration}. These results are consistent with the Brain Arousal model: moderate noise can improve cognitive performance, while excessive or insufficient noise are detrimental \cite{soderlund2008positive}. The findings support the potential for the improvement of network-security monitoring task performance through audio cues designed with appropriate levels of complexity.

While the potential utility of sonification for conveying network-security information is evidenced in prior work, and the integration of sonification into SOCs has been proposed, users' perspectives on this technology have not been explored. This is the research gap that our article seeks to address.

\section{Methodology}\label{sec:methodology}

\subsection{Research Approach}

We aimed to identify requirements in sonification design and integration, and contexts of use for sonification in SOCs, as part of the user-centred design process \cite{maguire2002user}. By contexts of use, we refer to the conditions under which sonification could be used in SOC work \cite{maguire2001context}. The stages of our study are shown in Figure \ref{fig:requirements_analysis}, in relation to the requirements analysis process. This requirements analysis approach is widely used in prior literature and described by Maguire et al. \cite{maguire2002user}.

\tikzstyle{textbox} = [rectangle, rounded corners, minimum width=0.8cm, minimum height=1.5cm,text centered, text width=1.5cm, draw=black]

\tikzstyle{textbox2} = [rectangle, rounded corners, minimum width=0.8cm, minimum height=1.5cm,text centered, text width=2.1cm, draw=black]
\tikzstyle{textbox1} = [rectangle, rounded corners, minimum width=0.8cm, minimum height=1.5cm,text centered, text width=4cm]
\tikzstyle{textbox3} = [rectangle, rounded corners, minimum width=0.8cm, minimum height=1.5cm,text centered, text width=7.4cm]

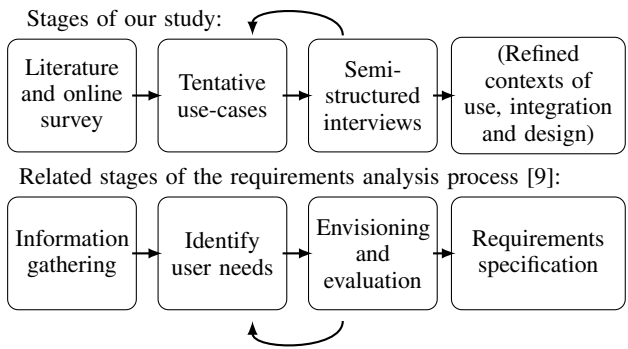
\begin{figure}[ht]
\centering
\small
\begin{tikzpicture}
\node (start) [textbox1] at (-5.3,1) {Stages of our study:};
\node (start) [textbox3] at (-3.1,-1.1) {Related stages of the requirements analysis process \cite{maguire2002user}:};
\node (start) [textbox] at (-6, -2.1) {Information gathering};
\node (start) [textbox] at (-4, -2.1) {Identify user needs};
\node (start) [textbox] at (-2,-2.1) {Envisioning and evaluation};
\node (start) [textbox2] at (0.2,-2.1) {Requirements specification};
\draw[->, >=latex, thick] (-5.2,0) -- (-4.8,0);
\draw[->, >=latex, thick] (-1.2,0) -- (-0.9,0);
\draw[->, >=latex, thick] (-5.2,-2.1) -- (-4.8,-2.1);
\draw[->, >=latex, thick] (-1.2,-2.1) -- (-0.9,-2.1);
\draw[->, >=latex, thick] (-3.2,0) -- (-2.8,0);
\draw[->, >=latex, thick] (-3.2,-2.1) -- (-2.8,-2.1);
\draw[->, >=latex, thick, bend right=90, distance=0.4cm] (-2.4, 0.8) to (-3.6,0.8);
\draw[->, >=latex, thick, bend left=90, distance=0.4cm] (-2.4, -3) to (-3.6,-3);
\node (start) [textbox] at (-6, 0) {Literature and online survey};
\node (start) [textbox] at (-4, 0) {Tentative use-cases};
\node (start) [textbox] at (-2,0) {Semi-structured interviews};
\node (start) [textbox2] at (0.2,0) {(Refined contexts of use, integration and design)};
\end{tikzpicture}
\caption{Study Methodology: Requirements Analysis Process}\label{fig:requirements_analysis}
\end{figure}

As illustrated in Figure \ref{fig:requirements_analysis}, we drew on existing literature and the results of an online survey to design tentative use-cases: descriptions of conditions in which sonification might be used in SOCs, the development of which is presented in Section \ref{sec:use-case_development}. We refined those use-cases that participants felt had some utility in the interviews, to produce contexts of use. 

By exploring the use-cases in interviews, we identified the potential for integrating sonification into SOCs, and challenges. Questions remain to be answered, however, before a full requirements specification (the final stage in Figure \ref{fig:requirements_analysis}) can be produced. The refined contexts of use, and integration and design requirements that we contribute in this paper are initial work that can form the basis of a requirements specification. In Section \ref{sec:discussion}, we highlight the areas that remain to be addressed experimentally and through further interaction with users in the construction of a full requirements specification.

To ensure face validity \cite{nevo1985face} of the online survey and interview questions, both were discussed with, and incorporated feedback from, a field expert (a researcher in HCI), and three subject matter experts (who worked, or had previously worked, in SOCs). Both the survey and interview questions were also answered by subject matter experts in a pilot study. 

We recruited a convenience sample of 20 participants for the online survey, and 21 participants for the interview. Participants were security practitioners who worked in SOCs with which we had previously established relationships, and were recruited through spoken or email contact with those responsible for the SOC. We targeted organisations that ran internal or multitenanted SOCs. There was likely some overlap between survey and interview participants, since the same SOCs were involved in each. The extent of this overlap is unknown, since survey responses were anonymised.

Ethical approval for this study was granted by the Central Univerity Research Ethics Committee, University of Oxford (reference: R48822/RE001). We ensured ethical handling of collected data through an informed consent process for participants, storage of data in password-protected files viewed only by the researchers, and anonymisation of published results.

\subsection{Developing Tentative Use-Cases}\label{sec:use-case_development}

We drew on existing literature in developing ideas for tentative use-cases for sonification in SOCs \cite{baier2007event, baldassi2006visual, ballora2011songs, ballora2012use, d2016cyber, etoty2014survey, hildebrandt2016continuous, merced2013sound, qi2007toward, sundaramurthy2014tale, van2009multisensory, zhang2012survey}. From this, we identified areas requiring validation, and constructed questions on these aspects. For example, one area in need of validation was the extent to which security practitioners were required to use multiple screens in SOCs, and for this we developed questions pertaining to the need to watch multiple monitors or dashboards. These questions were asked in an online survey of 20 security practitioners, in which participants were asked to indicate their level of agreement with 6 assertions. 

Level of agreement with assertions was indicated using a Likert-type scale \cite{likert1932technique} with response categories ``\textit{Strongly disagree}'' (=1) - ``\textit{Strongly agree}'' (=5). We selected the Likert-type scale as an efficient method of collecting participants' attitudes \cite{kaptein2010powerful}.  Based on these responses, we designed 5 tentative use-cases, which are presented in Section \ref{sec:use-case_development}.

\subsection{Semi-Structured Interviews} 

Face-to-face interviews took place at the organisations at which participants worked, in rooms exterior to the SOC. The exception was two participants who were interviewed through a live video chat due to travel constraints. Interviews were audio-recorded and lasted approximately 30 minutes. 

First, participants were introduced to the sonification prototype, a system that maps properties of network packets to music. The system reads a packet capture in (mock) real-time, and generates sound events based on sampled packets. Table \ref{table:prototype_mappings}, which we provide to enable replication of our research approach, describes the mappings used from properties of packets to musical properties. The prototype design is not the focus of this paper, so we do not detail the implementation. Further details on our technical approach can be found in \cite{axon2017formalised}.

\begin{table}[ht]
\caption{Sonification Prototype Mappings}
\label{table:prototype_mappings}
\centering
\small
\begin{tabular}{p{0.2\textwidth}p{0.2\textwidth}} 
\textbf{Packet property} & \textbf{Musical property} \\ 
\arrayrulecolor{black!25}\hline
IP/port common-ness & Consonance of pitch \\ 
Source/destination IP/port & Octave of pitch \\
Packet size & Amplitude \\
Direction of traffic & Pan of sound \\
\end{tabular}
\end{table}

The prototype was pre-recorded running on a synthetically-generated dataset containing port scan, DDoS, and data exfiltration attacks. Our aim here was to familiarise participants with the concept of sonification; this was particularly important given that the technique is relatively little known, and not operational in SOCs. Early prototyping is key to user-centred design, to convey to users an understanding of the system, elicit ideas for discussion, and enable users to play a role in the iterative design process \cite{gulliksen2003key}. This is also crucial for creating security interfaces that are effective, yet usable \cite{nurse2011guidelines}.  

The researchers described the system and mappings from data to sound. Participants then listened to an audio recording of the prototype using headphones. Next, the interview took place, guided by the questions presented below. We chose to conduct semi-structured interviews, with the aim of extending discussion based on the flow of conversation.

\begin{quote}

[1-5.] \textit{We are considering the use of sonification for [tentative Use-Cases 1--5] in SOCs. What is your view on the potential of sonification in this use-case? This can include this particular prototype, and also the concept of sonification as a whole for SOCs.}
\end{quote}

Before these questions were asked, participants were given the 5 use-cases on paper. Participants then answered each question, and discussion ensued with the researchers, expanding on topics brought up by the participant such as other use-cases, and challenges in integration. We encouraged both criticisms and positive responses. Throughout the interview, we highlighted that the participant could consider different sonification designs to the prototype presented. We ensured that this was clear, since the aim of the interview was to discuss the potential for the concept of sonification in SOCs in general.

Participants were then asked to rate the potential utility of each of the 5 tentative use-cases presented, using a Likert-type scale : ``\textit{Please rate the potential utility of sonification in this use-case, from 1: not at all useful, to 5: very useful}''. This rating stage was placed at the end of the discussion of each use-case to allow participants to formulate their views.

\subsection{Data Analysis}

Given discrepancy in the community as to how to treat Likert scale data \cite{jamieson2004likert, norman2010likert, robertson2012likert}, we calculated the mode and median to analyse both the responses to the assertions in the online survey, and the ratings given to each use-case in the interviews. We considered that a mode or median rating higher than 3 constituted overall agreement with an assertion, since 3 was the middle value.  We also calculated a comparison of non-neutral scores (CNNS), in which we took the ratio of scores less than (1, 2) and greater than (4, 5) the neutral value (3). The three measures support the same conclusions, considered alongside the analysis of the interview data. 

We analysed the interviews using template analysis \cite{king1998template}. This technique if useful for qualitative data analysis in which the researcher has some understanding of the concepts to be identified. We first developed a-priori themes to be identified in the data: use-case utility; integration questions; and design requirements. We manually transcribed our interview recordings, producing transcripts for each discussion, and spent time becoming familiarised with the data. We then coded the interview transcript dataset initially, attaching relevant parts of the transcriptions to the a-priori themes. Relevant sections of data that did not fit into these themes were assigned new codes. 

We thus produced an initial template of codes, which we then developed through iterative application to the dataset, modifying the template as appropriate to the data. Through this refinement we produced a final template and dataset coded according to it. We then interpreted the data and wrote up the findings within the themes of the template. During the interpretation and write-up process, we engaged in frequent reflections to avoid bias and the influence of personal beliefs.

\section{Development of Tentative Use-Cases}\label{sec:use-case_development} 

We summarise our development of ideas using existing literature on SOC working practice and sonification, indicating potential uses for sonification in SOCs. We present the outstanding questions (\textbf{OQ}s) that we identified and addressed to support the evolution of these ideas, and their formulation into assertions in an online survey. Finally, we present the 5 tentative use-cases derived. 

\subsection{Developing Ideas Using Existing Literature}

\begin{table*}[ht]
\caption{Online Survey Results: Responses to Assertions (Resp, ordered from ``\textit{Strongly disagree}'' (=1) -- \textit{``Strongly agree}'' (=5)): Mode, Median (Med), and Comparison of Non-Neutral Scores -- Disagree (1-2): Agree (4-5) (CNNS: D:A)}
\label{table:online_survey_results}
\centering
\small
\begin{tabular}{p{0.78\textwidth}|p{0.06\textwidth}|p{0.01\textwidth}|p{0.01\textwidth}|p{0.02\textwidth}} 
\textbf{Assertion} & \rot{\textbf{Resp}} &
\rot{\textbf{Mode}} & \rot{\textbf{Med}} & \rot{\textbf{CNNS}} \\ \hline
\textbf{Anomaly detection by humans} (pertains to \textbf{OQ1}) &&&&\\
\textit{Assertion 1}: Human analysts monitoring the network are capable of detecting network anomalies missed by automated systems                                                              & 0,1,5,10 & 4 & 4 & 1:14\\
\textit{Assertion 2}: The monitoring setup I use enables me to detect network anomalies that are missed by automated systems                                                                             & 0,4,11,4,1 & 3 & 3 & 4:5\\
\textit{Assertion 3}: I sometimes rely on my experience and intuition to detect network anomalies rather than monitoring system alerts                                                                   & 0,2,7,7,4 & 3.5 & 4 & 2:11\\
\textbf{Multitasking/non-primary task monitoring} (pertains to \textbf{OQ2})                                                                                                                                  &&&&\\
\textit{Assertion 4}: I am required to monitor the network, while carrying out other tasks simultaneously (e.g., responding to emails)                                                                    & 0,2,2,13,3 & 4 & 4 & 2:16 \\ 
\textbf{Monitoring across multiple screens} (pertains to \textbf{OQ3})                                                                                                                                        &&&&\\
\textit{Assertion 5}: In monitoring, I am required to watch multiple monitors depicting different data at one time                                                                                       & 0,1,2,12,5 & 4 & 4 & 1:17\\
\textit{Assertion 6}: I am required to watch multiple dashboards on the same monitor depicting data at one time                                                                                          & 0,3,4,9,4 & 4 & 4 & 3:13 \\
                                                                                                                                                                                 
\end{tabular}
\end{table*}

Anomaly-detection approaches for security monitoring are widely researched, including visualization-based techniques to enable detection of abnormal activity by humans \cite{etoty2014survey, zhang2012survey}. A wide array of experimental results evidence the utility of sonification for detecting anomalous patterns in data in fields including Medicine and Astrophysics, for example \cite{baier2007event, ballora2012use, merced2013sound, van2009multisensory}. Furthermore, prior work has supported the use of sonification for hearing network attacks \cite{ballora2011songs, qi2007toward}. We therefore posit that it is important to explore the potential for sonification to enable humans working in SOCs to detect anomalies in the network traffic, and seek to address the following question:

\begin{quote}
\textbf{OQ1.} Do security practitioners feel capable of detecting anomalies directly from the network traffic?
\end{quote}

Security practitioners may be required to carry out other tasks while monitoring the network; for example, managing email inboxes \cite{sundaramurthy2014tale}. Prior literature indicates the utility of sonification as a solution to enabling monitoring as a non-primary task. Hildebrandt et al. showed that using sonification to monitor a process as a secondary task while performing a different primary task had no significant effect on performance in either task \cite{hildebrandt2016continuous}. The use of sonification for peripheral monitoring may extend to cases in which security practitioners wish to continue to monitor whilst outside of the SOC. We consider that this may be true particularly for practitioners alone on shift, while taking breaks for example. To support the evolution of this idea, we seek to address the following question:

\begin{quote}
\textbf{OQ2.} To what extent are security practitioners required to multitask while monitoring in SOCs?
\end{quote}

The information required for monitoring in SOCs is often distributed across multiple monitors used by security practitioners \cite{d2016cyber}, including large screens at the front of the SOC. Security practitioners may therefore be required to focus their visual attention in multiple directions, yet it has been shown that visual perceptual clutter leads to increased errors in judgement \cite{baldassi2006visual}. Furthermore, security practitioners, depending on their role, can be required to monitor screens for extended time periods, focusing on visual representations of the data and monitoring alerts from SIEM solutions, for example \cite{sundaramurthy2014tale}, which may lead to visual fatigue. Presenting sonified data could reduce the emphasis on visual monitoring. This could mean either reducing the number of directions in which visual focus is required, or providing an alternative monitoring method for visually-fatigued practitioners. We seek to address the following question in developing this idea:

\begin{quote}
\textbf{OQ3.} To what extent are security practitioners required to visually monitor information presented on multiple screens?
\end{quote}

\subsection{Exploring Ideas Using an Online Survey}

The 6 assertions developed to assess the \textbf{OQ}s, and participants' responses to them, are presented in Table \ref{table:online_survey_results}. The online survey was completed by 20 participants working in SOCs: 2 SOC managers; 14 security analysts, 5 of whom were ``senior'' security analysts; and 4 (2 senior) security engineers. 

Five of the assertions obtained mode and median ratings greater than 3, which we consider agreement, as explained in Section \ref{sec:methodology}. The exception is \textit{Assertion 2}, which indicates that while practitioners feel capable of detecting anomalies, they are less confident that their existing monitoring setups enable this, and this is supported by the CNNS. This result supports experimentation with new methods of enabling this capability. The survey results can therefore be seen to affirm the 3 \textbf{OQ}s.

\subsection{Tentative Use-Cases}

Based on the survey results presented in Table \ref{table:online_survey_results}, and the prior literature, we derived the following 5 tentative use-cases to carry forward to the interviews.

\begin{enumerate}
\item \textbf{Detecting anomalies in the network traffic.} Presenting high-resolution sonifications of the network traffic, to enable humans to hear network anomalies.
\item \textbf{Monitoring as a non-primary task.} Sonifying network-security data to be monitored as a secondary task, enabling the user to carry out a separate primary task simultaneously.
\item \textbf{Monitoring data presented across multiple screens.} Sonifying parts of information that are currently presented across multiple screens, reducing the directions for focus of visual attention by users.
\item \textbf{Alleviating fatigue from monitoring screens.} Enabling users to monitor with reduced strain on visual attention, by providing the option to use sonification.
\item \textbf{Enabling monitoring whilst outside of the SOC.} Enabling users to continue monitoring work (e.g., using wireless earpieces) whilst outside of the SOC.
\end{enumerate}

Use-Case 1 was supported by the assertions of survey participants that detecting anomalies directly from the traffic was a capability of practitioners. The requirement to monitor across multiple screens motivated the development of Use-Case 3. Use-Case 4 was supported by the requirement for extended periods of visual monitoring reported in prior literature \cite{sundaramurthy2014tale}\hl. The requirement affirmed by the survey to multitask while monitoring the network justified the development of Use-Cases 2 and 5. We considered that multitasking might occur while carrying out other work inside the SOC, or while carrying out activities when away from the SOC, but still on duty.

\section{Interview Results}\label{sec:interview_results}

In this section we present demographics of interview participants. We report interview results relating to use-cases, integration, and design requirements, in Sections \ref{sec:use-case_interview} and \ref{sec:integration_interview}.

\subsection{Participants}

We interviewed 21 participants between May and June 2017. Participants were security practitioners working in 7 different SOCs. From 3 different internal SOCs, responsible for the security of a single organisation, 12 participants were interviewed. We also interviewed 9 participants from 4 different multitenanted SOCs, who provided managed services for client organisations. Of the participants, 4 were SOC managers; 10 were security analysts (3 senior); 2 were both security analyst and engineer; 2 were security engineers. Table \ref{table:participant_demographics} shows the job role and organisation type of each participant.

\begin{table}[h!]
\caption{Interview Participant (P) Demographics}
\label{table:participant_demographics}
\centering
\small
\begin{tabular}{p{0.14\textwidth}!{\color{black!25}{\vrule width 1pt}}p{0.14\textwidth}p{0.11\textwidth}}
                  & Internal SOC & Multitenanted SOC \\ \arrayrulecolor{black!25}\hline
Manager     & 3 (P1/2/17)          & 1 (P6)  \\ 
Senior Analyst & 0           & 3 (P7/15/16)   \\ 
Analyst & 7 (P3/4/13/18-21) & 3 (P10-12) \\ 
Engineer & 2 (P5/14)  & 0 \\ 
Analyst \& Engineer & 0 & 2 (P8/9) \\   
\end{tabular}
\end{table}

\section{Perspectives on Utility of Use-Cases}\label{sec:results_use-cases}\label{sec:use-case_interview}

We present the interview results on each tentative use-case, as well as new use-cases proposed by participants. Participants' views on the potential utility and challenges of each use-case are analysed, with the aim of refining promising contexts for the use of sonification in SOCs. In Section \ref{sec:discussion}, we reflect critically on the requirements for these contexts of use.

\subsection{Use-Case 1: Detecting Anomalies in the Network Traffic}

Overall, participants felt that sonification had potential in this use-case. A number of participants felt that humans were capable of detecting anomalies when presented with network data. The belief that it was mostly humans who detect network anomalies was expressed (P15), and it was suggested that humans have the capacity to recognise more subtle anomalies than machines: ``\textit{there's still a lot of human analysis, and a machine can only determine the really obvious ones}'' (P8). Security visualizations were frequently specified as a class of tool that enabled participants to detect anomalies, showing anomalous spikes in traffic volume, for example.

Possible benefits of sonification over existing anomaly-detection approaches were explored. The potential of sonification for detecting anomalies not apparent from visualizations was described, because ``\textit{the thing with a graph is -- it's not how much you can see, it's how much you can present}'' (P6). The trustworthiness of the information conveyed by the sonification was highlighted as an advantage over automated approaches, which can produce false-positives: ``\textit{it can't ever lie because it's just going on what it's seeing, it's not saying it's malicious it's just saying that that's what I am seeing}" (P10).

The ways in which anomalies might be detected using sonification were discussed; in particular, the potential to learn some baseline sound of the network, and from this basis detect anomalies. This included hearing deviations to greater traffic throughput. The potential to ``\textit{get used to the sound}'' such that deviations were apparent was also highlighted: ``\textit{when say a DoS attack or some other form of attack would take place, I'm sure it would stand out because you would get used to hearing a certain type of tune or hum from day-to-day activity}" (P15).

In general, participants felt that sonification had promise in this use-case. Assessment of the key points highlighted is required -- the ability to hear deviations from a ``baseline'' sound, and the comparison of a sonification-based approach to anomaly detection with automated and visualization-based approaches. This comparison is important given that, while many participants believed humans could detect anomalies in the data, some felt that anomaly-detection currently was predominantly machine-driven. Another participant noted that the real solution may be somewhere in between, i.e., that anomaly-detection capabilities differ between individuals (P16).

\subsection{Use-Case 2: Monitoring as a Non-Primary Task}

The general concensus was that sonification could prove valuable in this use-case. Participants stated that they were required to multitask in their role, and reported a range of tasks during which they were required to multitask whilst monitoring. These included researching new threats, composing reports, sending emails, or investigating cyber incidents:

\begin{quote}
\textit{One issue we have is that when we see something of interest, and we are researching [...] you're no longer monitoring. So, at points in time where you're not monitoring, if there was an audible cue that ``oh actually, there is something happening right now, maybe my attention should be back there''} (P13).
\end{quote}

The current requirement to use a visual means in multiple tasks was highlighted as a challenge: ``\textit{If we're investigating something else [...] I've only got three screens, and I've only got one pair of eyes}'' (P10). Participants described the potential value of sonification for monitoring without focused visual attention: ``\textit{you could just be monitoring or listening to that background rather than having to keep looking up}'' (P8). This extended to the use of sonification for monitoring alerts generated by automated systems, removing the need to keep ``\textit{viewing the alarm view while I'm doing other things}'' (P7).

The discussion of both sonified network traffic, and of auditory alerts, brings into question the types of information most appropriate for sonification in this multitasking application. The information content of sonified network packets, compared with auditory alerts, was highlighted as advantageous by one participant: ``\textit{the music can tell me, something else has happened [...] and not just as an -- alert, alert, alert}'' (P8).

In summary, perspectives on this use-case were positive, subject to some design and capability questions. A key question was the type of information to be sonified -- both network packets, and alert data, were discussed as advantageous. Participants voiced concerns about the possible effect of monitoring using sonification on their primary task, and vice versa. While Hildebrandt et al. showed that these effects were not significant in a different context \cite{hildebrandt2016continuous}, assessment with SOC-specific tasks, which are often time-pressured, and require high levels of attention, is required. The nature of SOC tasks could affect the performance of users multitasking using sonification.

\subsection{Use-Case 3: Monitoring Data Presented Across Multiple Screens}

The potential for sonification in this use-case divided opinion. Firstly, the extent to which practitioners felt that they were required to monitor across multiple screens differed between SOCs. Some (8/21) stated that multiple screens (between 2 and 7) were used to show alarms, devices, and chat feeds, for example. In other SOCs, all monitoring information was presented within a single pane of glass (6/21).  

Some challenges in the use of multiple screens were reported. Information could be missed because of its distribution across multiple screens. Missed information on monitors at the front of the SOC was reported, if practitioners were engaged by other screens: ``\textit{something on this} [front] \textit{screen could be red, but if they're already doing a priority 1, they're not going to look over there seeing the other priority 1}" (P6).

For these participants, monitoring across multiple screens was a challenge sonification could help alleviate. Both sonification of alerts and of network traffic were mentioned:

\begin{quote}
\textit{There are analysts sitting down there, and you have a massive dashboard, so they are still required to be looking at that at all times, and looking at their own screen. Sound will help in minimising that, just looking, as it avoids constant attention (P17)}.
\end{quote}

For some participants, however, the use of multiple screens did not pose a challenge, and it was considered convenient to have dedicated screens for executing commands, for example. These participants stated sonification would not be useful in this application, and did not wish to reduce the number of screens: ``\textit{I will still use 7 [screens], even if I have all the sound in the world}'' (P12). One participant reported that reducing the number of screens would cause inconvenience: ``\textit{If I don't have enough screens, I've got to constantly minimise, maximise, and copy this and go here and it can be very difficult}" (P7).

On the whole, participants were divided as to whether sonification had the potential to be useful in this use-case. The type of information that might usefully be represented by the sonification was unclear, and a number of participants did not desire any fewer screens. While it is clear that the spread of screen locations can cause information to be missed, it is likely that other technologies would be more effective solutions than sonification, meeting the needs of a greater proportion of security practitioners. Some participants suggested that the combination of this information into a single pane of glass would be a solution preferable to sonification in this instance.

\subsection{Use-Case 4: Alleviating Fatigue From Monitoring Screens}

In general, sonification was not perceived to have potential as a solution here. Some participants (6/21) stated that they were sometimes visually fatigued by their monitoring work in the SOC, yet others stated that they were not visually fatigued as they were accustomed to looking at screens. It was suggested that the extent to which fatigue was felt differed between individuals and types of role: ``\textit{nowadays I am doing stuff all the time, but there was a period when I was just staring at, I think it was, 3 different monitors at once}'' (P9).

Methods used for mitigating fatigue currently included encouraging workers to take regular breaks. Another approach adopted was automating as much as possible. Participants questioned the practicality of using sonification as an alternative for visually fatigued practitioners. If the sonification played only when practitioners were fatigued, their ability to interpret information from it might be limited:

\begin{quote}
\textit{I can see it as an alternative to visualization for when you get to a point when your eyes are tired [...] the thing is if you only switch it on when you get to that point, then I think you won't really understand what normal would be, so you would still need it on in the background to some extent} (P15).
\end{quote}

A number of participants felt that sonification would not be useful for them in this application. Visual fatigue was already prevented through other approaches (automation and regular breaks), such that participants were not (or were unaware that they were) fatigued by visual monitoring work, stating that they would continue to look at screens even with sonification. The utility of sonification in this use-case was questionable, and the ways in which it might work in practice unclear.

\subsection{Use-Case 5: Monitoring Whilst Outside of the SOC}

Participants generally felt sonification had strong potential in this use-case: ``\textit{if you were just going out and you pop a pair of headphones on or whatever and you can hear, something is going on, I can jump back in}'' (P10). Specific times that could necessitate monitoring whilst outside of the SOC included during fire alarms, and while making drinks, on break, or going to the shop. This was particularly true for participants who were required to work one-man shifts: ``\textit{today it's only me here, and I did have to leave to the shop earlier}'' (P11). 

It was noted that using sonification in this application could be particularly useful for practitioners alone on shift: ``\textit{the first ever job I did in a SOC I was the only person in the room. You could definitely say that would help with that one}'' (P9). Smaller SOCs, in which one-man shifts occurred, as well as companies running their own SOCs, were mentioned as situations in which this capability might be especially helpful: ``\textit{the guy running his own SOC, the SOC won't be his only task, he might be plumbing computers in the main office, and want to come back in if something big happens}" (P6). 

It was reported that there were existing approaches to monitoring whilst outside of the SOC. This included emails sent to cellphones, and a sonic alarm used when on break: ``\textit{when I leave, I unmute it, so that I can go and put my feet up, and then if there's an alarm I would come}'' (P11). The potential value of a more informative sonic approach (than the simple alarm currently used) was discussed by this participant: 

\begin{quote}
\textit{If we had a melody like yours representing that, and I knew what the melody was playing and what it was, then maybe I wouldn't have to come and look at it} [on-screen alerts]\textit{, because I would be like ok it's something normal for this time [...] with the current beep, we don't know until we actually log in} (P11).
\end{quote}

The placement of monitoring screens in the break room was another approach currently used to indicate to practitioners that they were required in the SOC. A number of participants discussed being waved at through the window by other SOC workers, to attract their attention while on break. This was particularly true for analysts with higher skill levels, required for specific events. Participants felt that sonification could be useful as an approach to informing practitioners on break that they are required in the SOC, played through speakers in break areas (e.g., the kitchen), or through an earphone worn while on break: ``\textit{they wouldn't need to rush back, keep checking, they could just go about their business and know `right, when I hear that sound, I need to take whatever action'}" (P7).

The desire to use sonification for monitoring outside of the SOC might differ. For example, one SOC manager was of the opinion that monitoring should not be continued whilst on break, as it would defeat the purpose of the break. In general, however, this use-case was considered a promising solution to actual challenges faced by security practitioners.

\subsection{Use-Case Ratings}

Table \ref{table:usecase_ratings} presents participants' ratings for each use-case. 

\begin{table}[h!]
\caption{Ratings Given to Use-Cases by Participants: (Resp, ordered from ``\textit{Not at all useful}'' (=1) -- \textit{``Very useful}'' (=5)): Mode, Median (Med), and Comparison of Non-Neutral Scores -- Not useful (1-2): Useful (4-5) (CNNS: N:U)}
\label{table:usecase_ratings}
\centering
\small
\begin{tabular}{l|l|l|l|l}
\textbf{Use-Case} & \rot{\textbf{Resp}} & \rot{\textbf{Mode}} & \rot{\textbf{Med}} & \rot{\textbf{CNNS}} \\ \hline
\textbf{1: Anomaly detection} & 1,1,7,7,5 & 3.5 & 4 & 2:12 \\
\textbf{2: Multitasking} & 1,4,4,4,8 & 5 & 4 & 5:12 \\
\textbf{3: Multiple screens} &  3,2,2,8,6 & 4 & 4 & 5:14 \\
\textbf{4: Visual fatigue} &  3,6,4,3,5 & 2 & 3 & 9:8 \\
\textbf{5: Outside-SOC activities} & 0,1,1,3,16 & 5 & 5 & 1:19 \\
\end{tabular}
\end{table}

Use-Cases 1, 2, 3 and 5 obtained mode and median ratings greater than 3, which we consider indicates overall agreement with potential utility (see Section \ref{sec:methodology}). The CNNS shows that these four uses-cases were rated above neutral by most participants. Based on these results, we selected Use-Cases 1, 2 and 5 to form the basis of our refined contexts of use, presented in Section \ref{sec:discussion}. Although Use-Case 3 also scored ratings greater than 3, we chose to omit it from the contexts of use, based on the qualitative interview analysis, from which we concluded that other solutions to the challenge of multiple screens may be more appropriate for SOCs.

\subsection{Other Use-Cases Suggested by Participants}\label{subsec:results_other_use-cases}

Aside from the use-cases we presented, other uses were suggested by participants, falling under the following themes.

\subsubsection{Occasional use} 

It was suggested that the sonification could be used to occasionally check the sound of the network: ``\textit{I might listen to it once an hour, and go `[...]it doesn't sound the same at 1 o'clock today as it did at 1 o'clock the last three days'}" (P6). Sonification could be played for the duration of particular events, which could be useful for conveying the length of events, since: ``\textit{sometimes looking at data you might not fully understand when it started and when it ended}'' (P11). Similarly, sonification could be played in the background particularly at times when high-severity incidents were being dealt with, to act as an indicator for SOC workers when a new incident may require their attention (P8).

\subsubsection{Hunting for Anomalies}

One participant suggested the use of sonification as a threat-hunting tool, for analysts required to search data for anomalies retrospectively: ``\textit{if I put that on for five minutes, and it sounds anomalous, then I know there's five minutes' worth of packets worth looking at. Otherwise I might spend an hour just looking at some packets with nothing particularly interesting in}" (P6).

The potential to listen to the sonification at increased speed (fast-forwarding), both for conducting audio reviews of data retrospectively (``\textit{if you've got an alarm or a period that you're interested in}''), and for real-time monitoring was discussed: 

\begin{quote}
\textit{If you had a baseline amount of traffic, you could go ``I'll listen to a minute of that, now I'm going to listen to a minute of what has just gone through the sensors", maybe accelerated, you will then start straight away going ``that doesn't sound right''} (P6).
\end{quote}

\subsubsection{Improving SOC workflow}

It was suggested that a continuous soundtrack could improve SOC workflow by making practitioners aware more efficiently of events that are relevant to them, without the need for others to escalate to them: 

\begin{quote}
\textit{At the minute, it relies on the first person who sees those events to recognise it's bad, to then escalate [...] if you heard lots of anomalies, the people who it would be eventually escalated to would instantly know that, and could maybe start on it earlier} (P8).
\end{quote}

A manager suggested that sonification could take over some of their alert-handling workload, by verbally presenting the queue of alerts and their severity ratings: ``\textit{an audio prompt would give me more time: [...] if it's not shouting numbers out, I don't need to look at the queue}" (P6).

\section{Perspectives on Integration and Design}\label{sec:integration_interview}

We present key themes identified relating to the integration of sonification into the SOC environment, and to sonification design. We consolidate these results in Section \ref{sec:discussion}, and highlight challenges and implications for system development.

\subsection{Headphones or Speakers?}

A number of participants discussed whether the sonification would be best played through speakers or personal headphones. Some participants highlighted potential problems with playing the sonification through speakers --  for example, that if the sonification was made the soundtrack to the SOC, practitioners who were not monitoring would still have to listen to it ``\textit{when they're trying to concentrate on doing something else}'' (P2).

Some participants, however, felt that headphones were not always a desirable solution, as wearing headphones could isolate practitioners, or hamper collaboration. Alternative solutions were suggested, including the use of a single earpiece rather than headphones, suggested by two different participants, to enable practitioners to continue to collaborate.

\subsection{Existing SOC Workflow and Soundscape}

Some participants focused on integrating sonification with necessary SOC workflow in an unobtrusive way, noting that it should not prevent ``\textit{people being able to talk about what's going on}'' (P10). A need to standardise responses to sounds heard was suggested: ``\textit{everything we do is based around a procedure, so I'm not sure how you would [...] get everyone to conform to, `when you hear this, you do this'}" (P7).

Participants described existing SOC soundscapes. In some SOCs there was currently a soundtrack, such as radio for the whole room. In others, there was no deliberate noise, with practitioners listening to music at times through headphones: ``\textit{we don't have any audio [...] Occasionally people use headphones to listen to music, and on the odd occasion we will put music on}" (P15). If integrated into SOCs sonification must work appropriately with this range of existing soundscapes.

\subsection{Complexity of Networks}

Participants discussed the difficulty of finding unusual behaviour in networks: ``\textit{the more complex your network is, the more difficulty you have working out what is unusual}" (P6). A suggested approach to dealing with large amounts of network traffic was filtering sound by particular IP addresses or assets. 

One participant highlighted the issue of network complexity in the multitenanted SOC they worked in: ``\textit{I think if it was for an internal SOC for a specific company that probably would work better. Here because we're a managed services provider, I think there would be too many things going on}" (P10). Further research into differences in required design solutions for different SOC types is needed, such as filtering sound to focus on single networks for multitenanted SOCs.

\subsection{Sonification of Alerts}

Sonification of alerts was mentioned by a number of participants (6/21), as an approach to communicating critical events, or alerting on particular systems: ``\textit{using this would benefit us, if only the DDoS mitigation stuff that we use, or a subset of alarms or devices alerts us to anomalies via sound}" (P7). It was suggested that sonified alerts could be layered with the sonified low-level network traffic: ``\textit{you could tell the system to play music not just based upon the packet captures but based upon outputs of other things, signatures, outputs of x, y, z. Then you can build up two layers of that, so you could listen to the underlying traffic as part of an incident}" (P8).

\subsection{Mitigating Fatigue}

A number of participants (8/21) stated that they felt they would be fatigued by continuous exposure to the sonification. The potential for occasional use of the sonification was discussed in the context of listening fatigue: ``\textit{I guess you could use it as and when, but I think if you put that on somebody's head for a day, I think you would struggle with that}" (P6).

The potential for the sonification to be unobtrusive unless required was highlighted: ``\textit{music you can switch off to, but equally the anomalies in there, your brain is going to pick up on them and go that's changed, that's different}'' (P8). Designing sonifications that are unobtrusive in this way is a potential approach to mitigating fatiguing effects.

\section{Discussion and Implications}\label{sec:discussion}

We reflect on the results presented, with a view to summarising our three main contributions, listed in Section \ref{sec:introduction}. We refine contexts of use, then consider the implications for sonification design and integration. This can guide interface designers in developing sonification systems for SOCs.

\subsection{Refined Contexts of Use}

Based on the interview results relating to use-case utility, we refine contexts of use, identifying the potential actors, key usage scenarios, and relevant SOC workflow factors \cite{maguire2002user}.

\subsubsection{Monitoring whilst outside of the SOC}

There are times when it is ideal for security practitioners to be able to continue their monitoring work whilst outside of the SOC. This is particularly true for smaller SOCs, in which workers undertake one-man night shifts. Such workers, who might leave the SOC for a short time (e.g., to make a drink) or for a longer time (e.g., to go to the shop) saw potential value in the use of sonification to enable their monitoring work to continue. In larger SOCs, listening to sonification in break areas could improve SOC workflow for more experienced practitioners, who might currently be called (e.g., waved at physically) back into the SOC by others when their expertise is needed.

This capability could be enabled through wireless earpieces worn by workers when venturing outside the SOC, or by speakers playing sonification in break areas. As well as the network-packet sonification approach, of which the prototype presented was an example, sonified alert streams were highlighted as information that could be monitored at times outside the SOC. Sonification designs that enable both packet and alert representation, individually or in combination, would therefore be appropriate. Monitoring accuracy and attention during out-of-SOC activities should be compared with inside-SOC capabilities, to support the development of this use.

\subsubsection{Detecting anomalies in network traffic}

Situations in which sonification of network traffic has potential value as an anomaly-detection approach include long-term, continuous listening to the sonification for real-time detection of deviations. To support this use, anomaly-detection capabilities using sonification should be compared with those using security visualizations and automated tools. Prior sonification work indicates that malicious network activity can be detected using sonification \cite{ballora2011songs, qi2007toward}, but does not make this comparison. 

Short-term anomaly-detection uses include occasional checking of the sonification -- e.g., once per hour -- to compare with previous times. Another promising short-term use is retrospective analysis. Practitioners tasked with hunting retrospectively through data for anomalies suggested sonifications of the data could enable location of interesting packets requiring closer inspection. Research is needed into approaches to enable users to link anomalous sounds heard to the relevant data (in a text or visual form). For such tasks, listening to sonification played at increased speeds could enable users to sift through data from extended time periods more quickly.

\subsubsection{Multitasking whilst monitoring as a non-primary task}

Sonification is potentially useful for aiding security practitioners in carrying out monitoring tasks while conducting other primary tasks. It is important to assess this capability experimentally; in particular, the effect of primary tasks on secondary-task sonification monitoring, and vice versa. Such work can draw on the aforementioned work of Hildebrandt, which showed that such monitoring had no significant effect on either primary or secondary task \cite{hildebrandt2016continuous}. However, context-specific assessment is important, using primary tasks relevant to SOCs: sending emails, writing threat intelligence reports, and investigating incidents were some tasks described.

\subsection{The Need for Flexiblity in Interface Design}

Some key differences in opinion were highlighted, with implications for sonification design. Participants differed in their opinion on whether the sonification should use headphones, speakers, or single earpieces, and whether continuous or occasional use would be most appropriate. It is clear that the different approaches may suit different users and scenarios. It is therefore appropriate for sonification designs to be flexible, depending on use-case and user preference. Playing the audio through all mediums discussed should all be viable (e.g., spatialisation of different sounds through different ears is unsuitable for single earpiece listening), and the sonification approach should support both continuous and occasional use.

The analysis highlights a difference in requirements between multitentanted and internal SOCs. A participant working in a multitenanted SOC described the potential difficulty of using sonification in that environment, with large amounts of data for many customers, compared with an internal SOC. Further research into differences in the required design solutions across different SOC types is necessary. A solution for multitenanted SOC environments might be the provision of tool features to filter sound by the single SOCs to be monitored.

The prototype design presented in this study initiates the participatory design process \cite{gulliksen2003key}. This should be iterative, and as such future design of sonification systems for this application can draw on the design requirements we identified. Consulting users in the development process is especially important given that the technology is not operational in SOCs.

\subsection{Challenges in Integrating Sonification into SOCs}

Some challenges in integrating sonification into SOCs emerged from the interview responses. Appropriate integration of sonification with the existing SOC soundscapes reported is key if the technology is to be unobtrusive to users. In SOCs where the soundscape is silent, headphones or a sonification design that is unobtrusive could be used. Equally, the existing soundscape may affect sonification listening: the sounds produced may be drowned out in noisy SOCs.

It was highlighted that sonification should not distract users in a way detrimental to SOC activity. Sonification systems should be designed with appropriate sound complexity for particular tasks, since complexity of auditory stimuli has been shown to affect cognitive performance \cite{soderlund2008positive}. Reducing cognitive load is a key consideration for creating usable security interfaces \cite{nurse2011guidelines}. Less complex sound is needed for non-primary tasks, since less complex background auditory stimuli have been shown to improve the performance of security-critical tasks \cite{berg2017exploration}. Mapping highly complex network data to low-complexity sounds will pose a challenge.

The copious amounts of complex data present on networks exacerbates the challenge of designing sonification systems suitable for the SOC environment, since it makes finding a baseline of ``normal" behaviour difficult. Concerns were voiced in interviews that sonification systems representing such data could become cacophonous, and tuning systems to some network baseline would take time. The need to train users to use these systems, and understand the sounds of the networks monitored such that abnormalities could be identified, was also discussed as a potential challenge. The time required for adequate training of users, and for tuning of systems to networks, is a key factor affecting the utility of the approach.

Listening to sonification for extended time periods may be fatiguing. Fatigue caused by previous sonification designs has been reported \cite{hermann2011sonification} and was highlighted as a potential pitfall by a large number of participants. In integrating sonification into SOCs, therefore, it is important to consider mitigating fatiguing effects. Kramer argued that developing aesthetic sonifications can reduce listener fatigue, and prior work in such aesthetic sonification can be drawn on \cite{kramer1993auditory}. Another approach to mitigating fatigue, to be assessed experimentally, is to enable personalisation of the sounds listened to.

\subsection{Study Limitations}

Owing to the nature of the semi-structured interviews we conducted, there was variation in the level of detail in which different participants discussed each question. Furthermore, this paper can report only those contexts of use, challenges in integration, and design requirements highlighted in this study. It is possible that others would emerge in conversation with other participants. Consolidation of these findings through further studies would ensure coverage of all requirements. 

The presentation and discussion of the technology with practitioners could have caused acquiescence bias, in which participants agreed with statements by default. To mitigate this, we encouraged discussion around criticisms as well as positive responses in the interviews, and explored challenges raised by participants pertaining to environmental factors such as the noisiness of the SOC, complexity of networks, and the distractions that could be caused by sonification.

\section{Conclusion and Future Work}\label{sec:conclusion}

Sonification has promise as an approach to improving security practitioners' working practices in SOCs, based on SOC workflow and challenges, and evidence of the benefits sonification can offer. Using an online survey and semi-structured interview responses from practitioners, we explored perspectives on the incorporation of sonification into SOCs. Our results show that security practitioners see high potential for the use of sonification in a range of use-cases; in particular, for peripheral monitoring -- while multitasking with other work tasks, or whilst outside of the SOC. Participants also saw value in using sonification for anomaly detection, in an approach similar to the existing visualization techniques used in SOCs.

We identified challenges in integration, and requirements for design, that should be addressed in future research. In order to be appropriate for a range of different SOC types, SOC soundscapes, and practitioners' job roles, sonification tools should be flexible in design. More specifically, sonification should be playable through a range of mediums, and suitable for a range of different types and lengths of use. Sonification of alerts was a key area highlighted for further design investigation, as well as approaches to mitigating listener fatigue.

As future work, we intend to address the design and integration questions highlighted in this study. We also intend to validate experimentally the capability of SOC practitioners to use sonification in our refined contexts of use, in comparison with other SOC tools. Experimentation with sonification in real SOC settings, and in realistically complex networks, will be key to assessing the utility of sonification for SOCs, and the effect of sonification on the SOC and vice versa.

\bibliographystyle{IEEEtran}
\bibliography{bibliosocs}

\end{document}